\documentclass[aps,prb,twocolumn,showpacs,groupedaddress]{revtex4}
\usepackage{graphicx}% Include figure files
\DeclareGraphicsExtensions{.jpg, .pdf, .png}

\begin{document}

\title{High pressure effects in fluorinated HgBa$_2$Ca$_2$Cu$_3$O$_{8+\delta}$}

\author{M. Monteverde}
\thanks{Scholarship of CONICET of Argentina. Also at Laboratorio de Bajas Temperaturas,
Departamento de F\'{\i}sica, FCEyN, UBA, Ciudad Universitaria, (C1428EHA)
Buenos Aires, Argentina.} \affiliation{CRTBT, CNRS, BP166, 38042 Grenoble Cedex
09, France}
\author{C. Acha}
\thanks{Also fellow of CONICET of Argentina.}
\email{acha@df.uba.ar} \affiliation{Laboratorio de Bajas Temperaturas,
Departamento de F\'{\i}sica, FCEyN, Universidad de Buenos Aires, Ciudad
Universitaria, (C1428EHA) Buenos Aires, Argentina}
\author{M. N\'u\~nez-Regueiro}
\affiliation{CRTBT, CNRS, BP166, 38042 Grenoble Cedex 09, France}
\author{D. A. Pavlov}
\author{K. A. Lokshin}
\author{S. N. Putilin}
\author{E. V. Antipov}
\affiliation{Department of Chemistry, Moscow State University, Moscow 119992,
Russia}

\date{\today}

\begin{abstract}

We have measured the pressure sensitivity of $T_c$ in fluorinated
HgBa$_2$Ca$_2$Cu$_3$O$_{8+\delta}$ (Hg-1223) ceramic samples with different F
contents, applying pressures up to 30 GPa. We obtained that $T_c$ increases
with increasing pressure, reaching different maximum values, depending on the
F doping level, and decreases for a further increase of pressure. A new high
$T_c$ record (166 K $\pm$ 1 K) was achieved by applying pressure (23 GPa) in a
fluorinated Hg-1223 sample near the optimum doping level. Our results show
that all our samples are at the optimal doping, and that fluorine incorporation
decreases the crystallographic $a$-parameter concomitantly increasing the
maximum attainable $T_c$. This effect reveals that the compression of the $a$
axes is one of the keys that controls the $T_c$ of high temperature
superconductors.

\end{abstract}


\maketitle

%para 2 columnas
%\begin{multicols}{2}

%\narrowtext

%\pagebreak

%\section{INTRODUCTION}

Since the discovery of high $T_c$ superconductors (HTSC) many efforts have
been devoted to understand the sensitivity of $T_c$ to the structural
parameters~\cite{Griessen89,Takahashi95,Nunez97,Antipov02}. This knowledge can
facilitate the determination of the mechanism beneath superconductivity and
can provide a way to increase the $T_c$ of these materials. There is now a
consensus, considering the correlation between the appearance of the
superconducting state with the structural characteristics of the material, that
the highest $T_c$'s can be reached if n=3 flat CuO$_2$ planes and small Cu-O
in-plane distances (dCu-O=a/2 for flat planes) can be achieved~\cite{Nunez97}.
To reach the maximum $T_c$, the doping level of the CuO$_2$ planes is also an
important factor to consider. It was already established~\cite{Presland91}
that:

\begin{equation} \label{eq:Tc(n)}
T_c(n)=T_c^{M} \bigl [1-\beta(n-n_{op})^2 \bigr ] ,
\end{equation}

\noindent where $\beta \sim 83$, $n$ and $n_{op}=0.16$ are the doping level
and the optimum doping level of the CuO$_2$ planes, respectively, and $T_c^M$
the maximum attainable $T_c$ for a variation of the doping level. High
pressure experiments contributed with significant results in this quest. They
have shown, particularly for the Hg-based cuprate
superconductors~\cite{Nunez93,Gao94,Cao95,Acha97,Acha98}, that $T_c$ increases
with increasing pressure, following a quadratic law which depends on the
doping level of the sample ($n$), on the pressure-induced charge transfer
($dn/dP$) and on an intrinsic factor ($dT_c^M/dP$). In this phenomenological
model, pressure increases linearly the doping level of the CuO$_2$ planes but
also increases $T_c^M$. It can be shown that~\cite{Nunez97}:

\begin{equation}
\label{eq:Tc(P)} T_c(n,P) =  T_c(n,0) + a P + b P^2 ,
\end{equation}

\noindent where $$ a = (\frac{dT_c^M}{dP}) + 2(n_{op}-n) \beta T_c^M
(\frac{dn}{dP}), $$ and  $$ b = - \beta T_c^M (\frac{dn}{dP})^2 .$$

The question is: what besides doping controls the value of $T_c$? The origin
of this intrinsic factor is still unclear~\cite{Evandro97,Jansen01}. In some
theoretical studies it is argued that $T_c^M$ is regulated by the
interlayer~\cite{Chakra04} or the intralayer~\cite{Pavarini01} coupling,
showing in both cases that materials with a larger distance between CuO$_2$
planes (i.e. a larger $c$ parameter) are generally those with higher values of
$T_c^M$. On the other hand, Marsiglio et al.~\cite{Marsiglio90} showed that
relevant changes on $T_c$ are obtained only when the in-plane Cu-O distances
are changed. There are also experimental
results~\cite{Acha98,Abakumov98,Locquet98,Nunez97} that confirm either point
of view, indicating that the structural relevant parameter is, in one hand,
the $c$ or, on the other, the $a$ lattice parameter.

The incorporation of fluorine into the structure of the Hg-1223
superconductor~\cite{Lokshin01} and the study of the pressure sensitivity of
its $T_c$ gave us an opportunity to contribute experimentally to this search.
Fluorine partially replaces the oxygen located in the (HgO$_{\delta}$ layer),
producing a reduction of the in-plane Cu-O distance, while the small CuO$_2$
plane buckling is maintained. As a consequence, the onset of the
superconducting transition determined by ac susceptibility ($T_c^{\chi ac}$)
increases from 134 K to 138 K for optimally oxygenated and optimally
fluorinated compounds, respectively.

In this paper we have studied the pressure dependence of the $T_c$ of the
fluorinated Hg-1223 compounds, with different fluorine contents. Our results
point out the reduction of the $a$-parameter as the key factor that controls
the value of $T_c^M$.

%\section{Experimental}

All the fluorinated ceramic Hg-1223 samples (Hg-1223F) studied here were
synthesized and characterized previously~\cite{Lokshin01}. Resistivity as a
function of temperature (4 K $\leq T \leq$ 300 K) for pressures from 4 GPa to
approximately 30 GPa was measured for samples with different fluorine contents
labelled \#1 (a=3.8496 \AA, $T_c^{(\chi_{ac})}\simeq$ 138 K) and \#2 (a=3.8536
\AA, $T_c^{(\chi_{ac})}\simeq$ 135 K). The high pressure was applied using a
quasi-hydrostatic experimental setup, corresponding to a Bridgman
configuration with sintered diamond anvils, where pyrophillite is used as a
gasket and steatite as the pressure medium that favors quasi-hydrostatic
conditions. The superconducting transition of Lead is used to determine the
pressure inside the cell. The pressure gradient was estimated from the width
of this transition and corresponds to a 5-10\% of the applied pressure, for
pressures lower than 10 GPa, with a saturation's value of 1 GPa for higher
pressures and up to 30 GPa. A conventional 4 terminal DC technique was used to
measure resistivity under high pressure at different temperatures. Electrical
contacts were made using thin Pt wires pressed to the sample's surface by the
pressure setup. A well calibrated Cernox thermometer thermally anchored to the
anvils ensures a determination of sample's temperature with an uncertainty
lower than 0.2 K for the whole Tc range studied.

%\section{RESULTS}

\begin{figure}
\begin{center}
\includegraphics[width=3.5in]{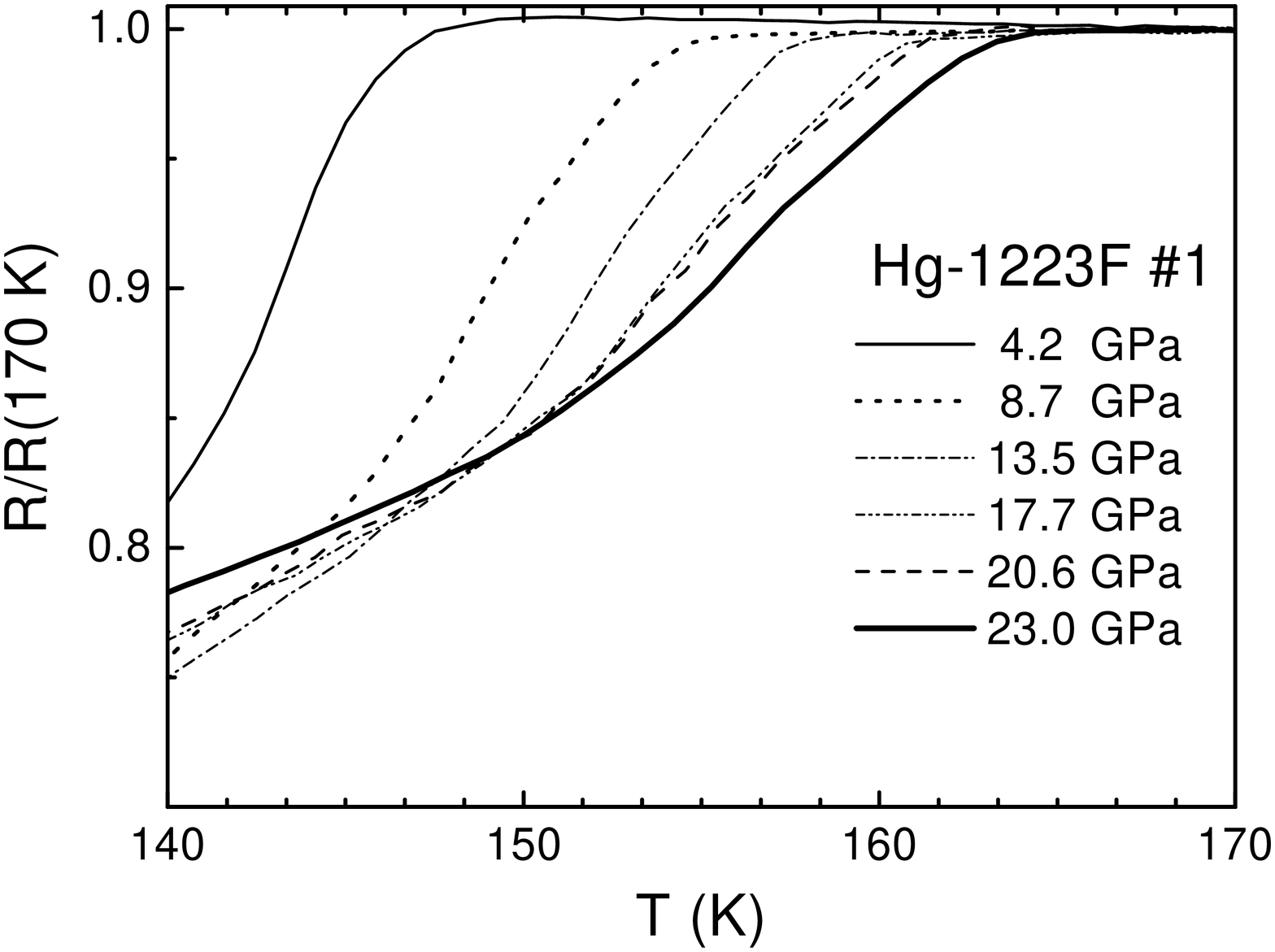}
\vspace{-5mm}
%\centerline{\epsfig{figure=renum.eps,width=7in}} \vspace{-15 mm}
\caption{Normalized resistance of sample \#1 as a function of temperature for
different applied pressures.} \vspace{1mm} \label{fig:RvsT}
\end{center}
\end{figure}

\begin{figure}
\begin{center}
\includegraphics[width=3.5in]{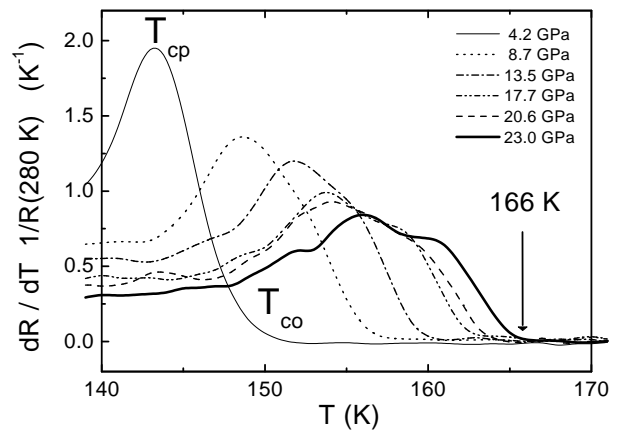}
\vspace{-5mm}
%\centerline{\epsfig{figure=renum.eps,width=7in}} \vspace{-15 mm}
\caption{Temperature derivative of the resistance of sample \#1 at different
pressures.$T_{cp}$ and $T_{co}$ illustrate how the peak and the onset
transition temperature are defined for $P$ = 4.2 GPa, respectively.}
\vspace{1mm} \label{fig:deriv}
\end{center}
\end{figure}

The effect of pressure on the resistance of sample \#1 can be observed in
Fig.~\ref{fig:RvsT}. A similar behavior was obtained for samples with other F
content. Zero resistance is achieved at low temperatures, in the range of
10-60 K depending on the quality of the intergrain coupling of these ceramic
samples. By using the temperature derivative of the resistance we can define
the onset critical temperature ($T_{co}$), where the derivative departs from
its normal behavior, and a peak transition temperature ($T_{cp}$) determined
by the peak of the derivative, as can be observed in Fig.~\ref{fig:deriv}. The
former criterion is usually dominated by thermal fluctuations and corresponds
to the formation of small superconducting droplets. This criterion was also
used in Ref.~\onlinecite{Gao94}. The latter is mostly related to the
appearance of a bulk superconductivity and gives a numeric value similar to
$T_c^{(\chi_{ac})}$. For the whole pressure range studied, the sample \#1
shows a $T_{co}$ higher than the one reported for the non-Fluorinated Hg-1223
samples.  In particular, at 23 GPa a $T_{co}$ of (166 $\pm$ 1) K is obtained,
which to our knowledge, is the highest ever reported.

\begin{figure}
\begin{center}
\includegraphics[height=3in,width=3.5in]{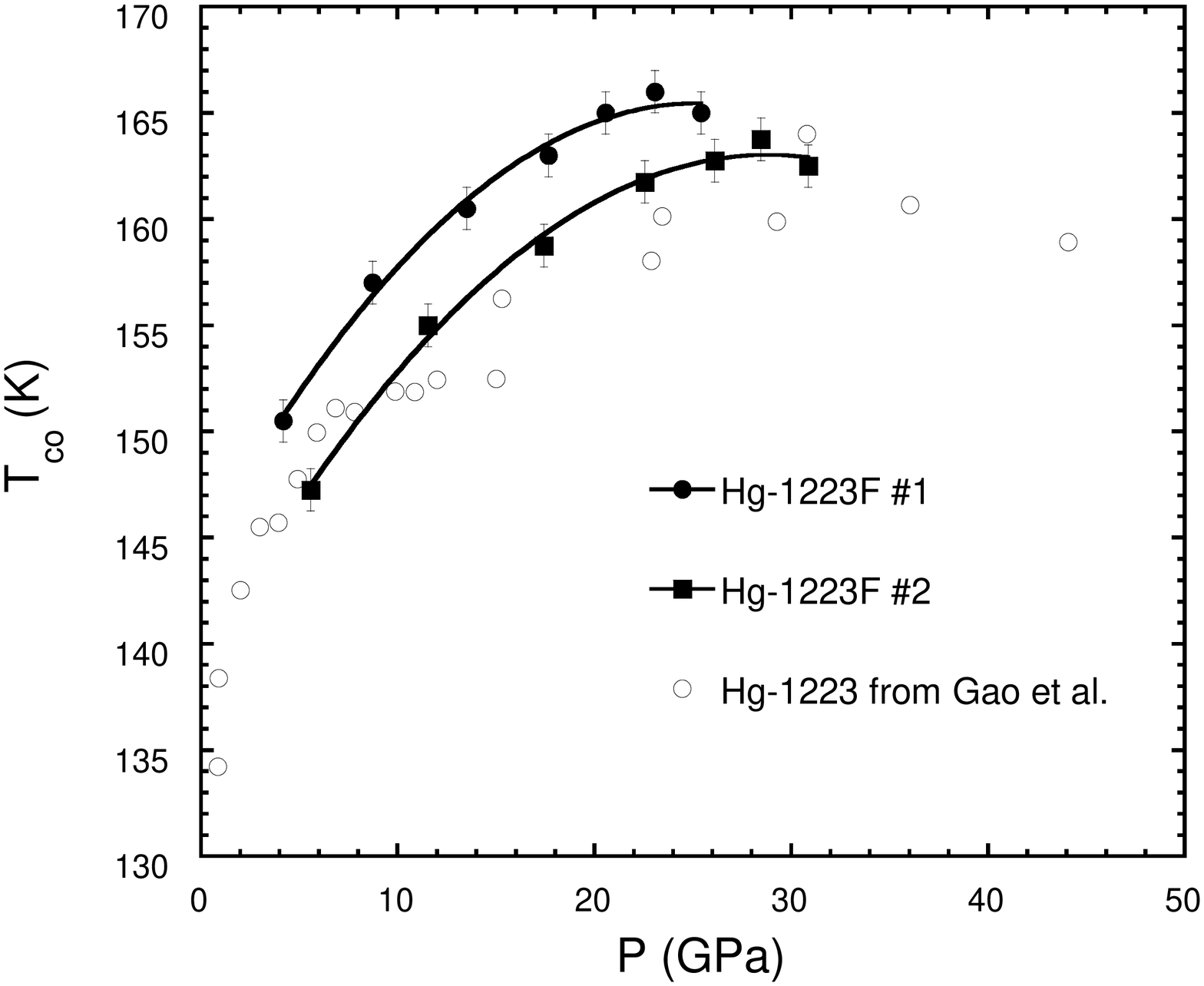}
\vspace{-5mm}
%\centerline{\epsfig{figure=renum.eps,width=7in}} \vspace{-15 mm}
\caption{Pressure dependence of $T_{co}$ for samples \#1 and \#2. For
comparison, the data of reference \onlinecite{Gao94} is also included. Lines
are fits using Eq. (\ref{eq:Tc(P)}).}

\vspace{1mm} \label{fig:TcovsP}
\end{center}
\end{figure}

As it is shown in Fig.~\ref{fig:TcovsP}, $T_{co}$ follows a parabolic
dependence with pressure which can be well described with Eq. (\ref{eq:Tc(P)}).

%\section{DISCUSSION}

All the Hg-1223F samples studied here under high pressure have a $T_c$ near
the flat maximum observed in the dependence of $T_c$ vs. the $a$-parameter (see
Fig.~3 of reference~\onlinecite{Lokshin01}). Thus, in principle, these samples
are near or at the optimum doping level. If we assume that sample \#1 is
optimally doped ($T_c^{\#1} = T_c^M$ = 138 K; $n^{\#1} = n_{op}$ = 0.16) then,
following Eq. (\ref{eq:Tc(n)}), sample \#2 ($T_c^{\#2} < T_c^M$) should have a
different doping level ($\mid n^{\#2}-n^{\#1} \mid \gtrsim$ 0.02
holes/CuO$_2$). Contrary to this, the fits of the $T_c(P)$ dependence using
Eq. (\ref{eq:Tc(P)}) indicate that both fluorinated samples have nearly the
same linear coefficient ($\sim$ 1.7 $\pm$ 0.1 K/GPa). If the pressure-induced
charge transfer of these samples is similar to the one reported for the
Hg-1223 system, this fact is indicating, according to Eq. (\ref{eq:Tc(P)}),
that both samples have an optimum doping level ($n=n_{op} \pm$ 0.005
holes/CuO$_2$). Indeed, the best fits for the $T_c(P)$ curves of samples \#1
and \#2 give a pressure-induced charge transfer coefficient very similar for
both samples [$(dn/dP)^{\#1} = $(1.7 $\pm$ 0.1) 10$^{-3}$ holes/GPa and
$(dn/dP)^{\#2} = $(1.6 $\pm$ 0.1) 10$^{-3}$ holes/GPa] that, as we supposed,
is near the value obtained for the optimally oxygenated
Hg-1223~\cite{Singh94b}. The same $dT_c^M/dP$ (= 1.7 $\pm$ 0.1  K/GPa) is also
obtained, but a different $T_c^M$ value for both samples. According to these
results, F incorporation for these samples is modifying essentially their
$T_c^M$ while keeping the doping of the CuO$_2$ planes in the optimum doping.
In other words, F is varying one of the structural parameters that controls
intrinsically the value of $T_c$. F reduces the structural-disorder as the
anion occupancy is higher in the fluorinated than in the oxygenated samples,
while the doping level is kept low, probably as a consequence of the charge
difference between F and O. As the main structural contribution of fluorine
incorporation is the reduction of the $a$-parameter, we may conclude that the
decrease of the $a$-parameter is one of the fundamental keys to the increase of
$T_c^M$.

The high values of $T_c$ obtained under pressure on the Hg-1223F samples are
probably the consequence of having a high intrinsic term ($dT_c^M/dP$) and a
small pressure-induced charge transfer which prevents a rapid overdoping of
the CuO$_2$ planes, even for samples optimally doped. The intrinsic term can
be associated with a positive contribution which comes from the reduction of
the $a$-parameter.

\begin{figure} [t]
{\includegraphics[height=3.2in,width=3.3in,angle=-0]{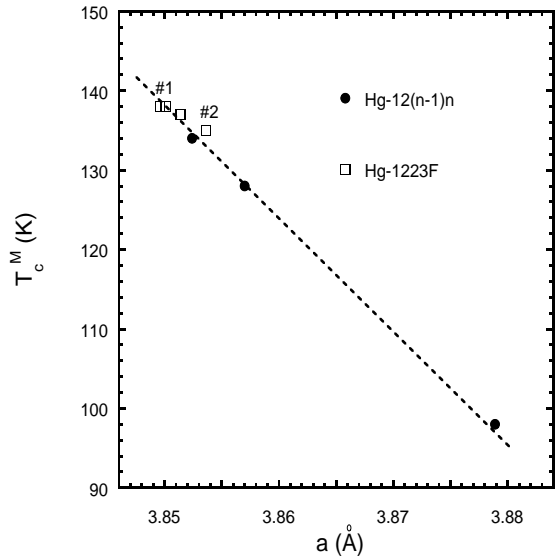}}
\vspace{-0mm} \caption{Structural sensitivity of the maximum critical
temperature $T_c^M$ on the chemical variation of the $a$ crystallographic
parameter, determined by X-Ray diffraction refinements. Intermediate points
for other Hg-1223F samples than sample \#1 and sample \#2 are also included.
The dashed line is a guide to the eye.}

\vspace{5mm} \label{fig:Tcvsa}
\end{figure}

The dependence of $T_c^M$ for the Hg-12(n-1)n series to chemical variations of
the $a$-parameter can be observed in Fig.~\ref{fig:Tcvsa}, where fluorine
incorporation has gradually extended this curve~\cite{Lokshin01} to lower
values of $a$. It should be noted, indeed, that the $T_c(a)$ dependence for
the Hg-1223F samples, near or in the optimum doping, seems to follow the
linear dependence of the $T_c^M(a)$ curve for the whole Hg series.

This means that samples \#1 and \#2 have practically the same doping level,
but a different $a$-parameter. We may track their differences to the way in
which the samples were prepared, as when synthesized, oxygen from the Hg-O
layer was extracted as much as possible and then fluorine was then
incorporated. The same doping with a different $a$ may be due to a different
F-O relation; the sample with smaller $a$ having a higher F to O ratio. The
structural refinements introduced by fluorine incorporation into the Hg-1223
structure can then provide valuable data in order to assess the structural
sensitivity of the doping mechanism and the origin of the intrinsic dependence
of $T_c^M$.

A simple determination of the pressure sensitivity of $T_c^M$ can be performed
using the data from the slope of the curve represented in Fig.~\ref{fig:Tcvsa}
and from the pressure dependence of the structural
parameters~\cite{Armstrong95}. Hence, for pressures up to 10 GPa, we determine
that $\frac{dT_c^M}{dP} = \frac{dT_c^M}{da} \frac{d(a)}{dP} \sim$ 10 K/GPa,
which overestimates the experimental value~\cite{Acha98} of $\sim$ 2 K/GPa. It
is clear that the variations of the $a$-parameter cannot fix solely the value
of $T_c^M$. A negative contribution of an additional parameter should be
considered, which can be possibly related to the increase of the buckling of
the CuO$_2$ layers~\cite{Lokshin01,Jorgensen95}. The small increase of
d$n$/d$P$ for sample \#1, easily noticed by the fact that a lower pressure is
needed to reach the maximum $T_c$, may indicate the proximity of a sudden
change of this parameter for further doping, as was observed from optimally to
highly oxygenated Hg-1201 samples~\cite{Cao95}.

Therefore, a large pressure-induced overdoping of the CuO$_2$ planes can be
predicted for the Hg-1223F samples with a lower $a$-parameter than that of
samples \#1 and \#2 ($a \leq$ 3.8496 $\AA$). This is indeed what was observed
when a further chemical compression was applied by increasing the fluorine
content in the Hg-1223F structure~\cite{Lokshin01}. The overdoping and the
chemical difficulties to produce small  $a$-parameters  without increasing the
buckling of the CuO$_2$ planes should be overcome in order to obtain higher
T$_c$'s than those obtained for the Hg-1223F compound under pressure.

To summarize, we have studied the pressure dependence of $T_c$ for the
Hg-1223F compound. In an optimally fluorine-doped sample we have obtained the
highest $T_c$ ever measured up to now. At an approximately constant doping
concentration, the optimal one, as neatly determined by our pressure
experiments, the $T_c^M$ increases with decreasing the $a$-parameter as a
consequence of the variation of the fluorine-oxygen ratio. This implies that
$a$ plays a major role on the determination of the superconducting state of
the HTSC. Further experimental results would be needed to clarify this issue,
determining if effectively uniaxial compressions along the $c$ axis would
produce minor effects on $T_c^M$.

This work was partially supported by a CONICET (Argentina) - CNRS (France)
cooperation grant. We are indebted to P. Tamborenea for a critical reading of
the manuscript.

% Create the reference section using BibTeX:
%\bibliographystyle{prsty}
%\bibliography{bibHgF}

\end{document}